\documentclass[aps,pre,twocolumn,groupedaddress,showpacs]{revtex4}
\usepackage{graphicx}
\usepackage{amsmath,amssymb,amsfonts}
\usepackage{natbib}

\newcommand{\bd}{\begin{displaymath}}
\newcommand{\ed}{\end{displaymath}}
\newcommand{\be}{\begin{equation}}
\newcommand{\ee}{\end{equation}}
\newcommand{\bs}{\begin{subequations}}
\newcommand{\es}{\end{subequations}}
\newcommand{\ba}{\begin{eqnarray}}
\newcommand{\ea}{\end{eqnarray}}

\begin{document}

\title{Surface diffusion and low vibrational motion with interacting
adsorbates:\\ A shot noise description}

\author{R. Mart\'{\i}nez-Casado$^{a,c}$}
\email{ruth@imaff.cfmac.csic.es}

\author{J. L. Vega$^{b,c}$}
\email{jlvega@imaff.cfmac.csic.es}

\author{A. S. Sanz$^c$}
\email{asanz@imaff.cfmac.csic.es}

\author{S. Miret-Art\'es$^c$}
\email{s.miret@imaff.cfmac.csic.es}

\affiliation{$^a$Lehrstuhl f\"ur Physikalische Chemie I,
Ruhr-Universit\"at Bochum, D-44801 Bochum, Germany}

\affiliation{$^b$Biosystems Group, School of Computing,
University of Leeds, Leeds LS2 9JT, United Kingdom}

\affiliation{$^b$Instituto de Matem\'aticas y F\'{\i}sica Fundamental,
Consejo Superior de Investigaciones Cient\'{\i}ficas,
Serrano 123, 28006 Madrid, Spain}

\date{\today}

\begin{abstract}
Here, an approach in terms of shot noise is proposed
to study and characterize surface diffusion and low vibrational
motion when having interacting adsorbates on surfaces.
In what we call {\it statistical limit}, that is, at long times and
high number of collisions, one expects that diffusing particles
display an essential Markovian behavior.
Accordingly, the action of the pairwise potentials accounting for
particle-particle collisions is equivalent to considering a shot
noise acting on a single particle.
We call this approach the {\it interacting single adsorbate
approximation}, which gathers three important advantages: (i) the
dynamics underlying surface diffusion and low vibrational motion
can be easily understood in terms of relatively simple stochastic
processes; (ii) from our model, appropriate (and well justified)
working formulas are easily obtained, which explain the results
arising from more complicated (but commonly used) molecular dynamics
simulations within the Langevin formulation; and (iii), at the same
time, it is less demanding
computationally than the latter type of calculations.
In order to illustrate the application of this model, numerical
results are presented.
Specially, our model reproduces the experimental observation regarding
the broadening of the quasielastic peak ruling surface diffusion.
\end{abstract}



\maketitle


\section{Introduction}
\label{sec1}

Diffusion of adsorbates (e.g., atoms or small molecules) on metal
surfaces is one of the most fundamental processes in surface science
for which there is a wealth of available experimental data.
These data are obtained by means of either standard time-of-flight
techniques \cite{toennies1,toennies2,toennies3} or the more
novel spin-echo measurements \cite{jardine1}.
In both cases the theoretical approach followed to understand and
explain the experimental results is basically the same.
At low coverages \cite{coverage} adsorbate-adsorbate interactions
can be neglected; diffusion ({\it self-diffusion}) is well
characterized by only studying the dynamics of a single adsorbate
interacting with the surface.
This approach is known as the {\it single adsorbate approximation},
where adsorbate dynamics are described by means of the (classical)
standard Langevin equation, i.e., the diffusion process is considered
as a Brownian-like motion \cite{toennies1,toennies2,JLvega1,sancho}.
On the other hand, as the coverage increases, adsorbate-adsorbate
interactions can no longer be neglected.
In these cases, molecular dynamics techniques within the Langevin
framework (MDL) are commonly used to study the problem to deal with
the surface thermal vibrations.
In this type of calculations the lateral dipole-dipole interaction
between Na atoms as a function of the coverage is accounted for by
pairwise potential functions.
Previous works \cite{ying,toennies3} using this approach have failed
when trying to reproduce the experimental data quantitatively.
However, recently it has been shown \cite{jardine2} that a good
agreement with the experiment can be achieved by considering the
adsorbate motion in three dimensions.

Introducing pairwise potential functions results in a realistic
description of the adsorbate dynamics.
However, there is not a simple manner to handle the resulting
calculations by means of a theoretical model, and one can only
proceed by using some suggested fitting functions.
Moreover, employing MDL simulations always result in a relatively high
computational cost due to the time spent by the codes in the
evaluation of the forces among particles.
This problem becomes worse when working with long-range interactions,
since {\it a priori} they imply that one should consider a relatively
large number of particles in order to obtain a good simulation.
In such cases, one can also make use of truncations, as is commonly
done when considering short-range interactions, but they might lead
to serious inaccuracies \cite{frenkel}.
Some alternative techniques have been proposed to overcome this
drawback, though they are more expensive computationally.

In order to avoid such inconvenient (interpretative and computational)
one can try to translate the behaviors observed in the experiment into
a simple, realistic stochastic model.
To achieve this goal, first note that when a few adsorbates are present
on a surface, the single adsorbate approximation perfectly describes
their dynamics \cite{note}.
For practical purposes, this means that long-range interactions can
be assumed as negligible; otherwise, important effects would appear
even when having adsorbates located at very far distances from each
other.
Hence, in principle, when having a higher coverage, one can assume
that one adsorbate will basically feel the action of other adsorbates
located in a certain neighborhood (as is corroborated by the
calculations presented in Ref.~\cite{jardine2}, where only
about 400 atoms were used in the simulations).
In other words, the importance of the long-range tail of the
particle-particle interaction becomes only relative.
Now, in order to obtain a good simulation of a diffusion process, one
has to consider very long times in comparison to the time scales
associated to the friction caused by the surface or to the typical
vibrational frequencies observed when the adsorbates keep moving
inside a surface well.
For example, in the Na/Cu(001) system the time scales associated to
typical vibrational frequencies and frictions are of the order of
a few picoseconds, while the propagation time employed in a typical
MDL simulation is of the order of several nanoseconds \cite{jardine2}.
This means that there will be a considerably large number of collisions
during the time elapsed in the propagation, and therefore that at some
point the past history of the adsorbate will be irrelevant.
That is, adsorbates undergo so many collisions after some time
that they lose any trace of the type of interaction among them when
they are considered statistically.
This memory loss is a signature of a Markovian dynamical regime, where
adsorbates have reached what we call the {\it statistical limit}.
Otherwise, for time scales relatively short, the interaction is not
Markovian and it is very important to take into account the effects
of the interactions on the particle and its dynamics (memory effects).

Based on the previous physical considerations, here we propose a
numerical and analytical scheme to deal with the diffusion problem
of interacting adsorbates as a function of the coverage.
In particular, our scheme is framed within a Langevin formulation,
and inspired by the theory of spectral-line collisional broadening
developed by Van Vleck and Weisskopf \cite{vvleck-weiss} and the
elementary kinetic theory of gases \cite{mcquarrie}.
We call this approach the {\it interacting single adsorbate
approximation}.
The diffusion of a single adsorbate is modeled by a series of
random pulses within a Markovian regime (i.e., pulses of relatively
short duration in comparison with the system relaxation) simulating
collisions between adsorbates.
In particular, we describe these adsorbate-adsorbate collisions by
means of a {\it white shot noise} \cite{gardiner} as a limiting case
of colored shot noise.
In this way, a typical MDL problem involving $N$ adsorbates is
substituted by the dynamics of a single adsorbate, where the action
of the remaining $N-1$ adparticles is replaced by a random force
given by the shot noise.
This type of noise has also been applied, for example, to study thermal
ratchets \cite{hanggi1} and mean first passage times \cite{iturbe}.
A different type of collisional model has been  recently used to obtain
jump distributions \cite{tommei}.
A general account on colored noise can be found in \cite{hanggi2}.

Note that since our model faces the problem of adsorbate diffusion from
a stochastic perspective, it is alternative to those other models that
imply the use of MDL simulations.
There are no incompatibilities between both approaches, stochastic
and deterministic (``deterministic'' in the sense of how
adsorbate-adsorbate interactions are handled), other than their
respective ranges of validity.
On the contrary, they complement each other, since the type of results
rendered by our model could be used to better understand the behaviors
and trends displayed by the fitting functions that appear in the
latter.

The main goal in this work is to show that a lot of information about
diffusion can be obtained by means of a simple stochastic model based
on a shot noise.
Thus, no substrate friction will be considered in our calculations.
The combined action of two noises, Gaussian white noise to simulate
the surface temperature and the white shot noise with a nonseparable
interaction potential are tackled in a separate publication
\cite{prb} (H\"anggi and co-workers \cite{hanggi1} have also used this
combined scheme in the context of Brownian motors).
In Sec.~\ref{sec2} we thus give a detailed account of the model that
we propose here to simulate the interacting particle dynamics.
Moreover, a discussion on the validity of the model is also provided
in the light of the {\it fluctuation-dissipation theorem}.
A brief account on the broadening of the quasielastic peak with
coverage is provided in Sec.~\ref{sec3}.
Numerical results for both flat and corrugated periodic surfaces
(separable interaction potentials) are shown in Sec.~\ref{sec4} as
a function of the coverage, temperature, and $\Delta K$.
Finally, the main conclusions extracted from this work as well
as a discussion on the range of validity of our model are given
in Sec.~\ref{sec5}.


\section{Interacting adsorbates and shot noise}
\label{sec2}

Particle motion under two-dimensional (2D) separable interaction
potentials can be simplified as two independent 1D motions provided
the correlations between the two components of the noise source along
each direction can be neglected.
Then, the motion of an adsorbate subjected to the action of a bath
consisting of another adsorbates on a static 2D (separable) surface
potential can be well described by a generalized Langevin equation
\be
 \ddot{x}(t) = - \int_0^t \gamma (t-t') \ \! \dot{x}(t') \ \! dt'
  + F[x(t)] + \delta R(t) ,
 \label{eq-langg}
\ee
where $x$ represents any of the two adsorbate degrees of freedom.
In Eq.~(\ref{eq-langg}), $\gamma(t)$ is the bath memory function;
$F = - \nabla V$ is the deterministic force, \mbox{$V(x) = V(x+a)$}
being a deterministic, phenomenological adiabatic potential accounting
for the adsorbate-surface interaction at $T=0$, with period $a$
along the $x$ direction; and $\delta R(t)$ is the stochastic force
fluctuation, defined as
\be
 \delta R(t) \equiv R(t) - \langle \langle R \rangle \rangle ,
 \label{rand-noise2}
\ee
with
\be
 \langle \langle R \rangle \rangle \equiv
  \sum_K P_K (\mathcal{T}) \langle R(t') \rangle_\mathcal{T} .
 \label{doubleav}
\ee
In this last expression, the double average bracket indicates the
average over both number of collisions ($K$) according to a certain
distribution, $P_K$, and time ($\mathcal{T}$); the subscript
$\mathcal{T}$ denotes the average along the time interval
$\mathcal{T}$.
Note that both $F$ and $\delta R$ are forces per mass unit.

The random force $R(t)$ has the functional form of a shot
noise,
\be
 R (t) = \sum_{k=1}^K b_k (t - t_k) ,
 \label{elect}
\ee
since it describes the collisions among the adsorbates.
The information about the shape and effective duration of the
$k$th adsorbate-adsorbate collision at $t_k$ is thus provided
by $b_k(t-t_k)$.
According to the definition of a shot noise, the probability to
observe $K$ collisions after a time $\mathcal{T}$ follows a Poisson
distribution \cite{gardiner}, given by
\be
 P_K (\mathcal{T}) =
  \frac{\ (\lambda \mathcal{T})^K}{K!} \ \! e^{-\lambda\mathcal{T}} ,
 \label{poisson}
\ee
where $\lambda$ is the average number of collisions per time unit.
Assuming sudden adsorbate-adsorbate collisions (i.e., strong but
elastic collisions) and that after-collision effects relax
exponentially at a constant rate $\lambda'$, the pulses in
Eq.~(\ref{elect}) can be modeled as
\be
 b_k(t-t_k) = c_k \lambda' {\rm e}^{- \lambda' (t-t_k)} ,
 \label{pulse}
\ee
with $t-t_k > 0$ and $c_k$ giving the intensity of the collision
impact.
Within a realistic model, collisions take place randomly at different
orientations and energies.
Hence it is reasonable to assume that the $c_k$ coefficients follow
an exponential law,
\be
 g(c_k) = \frac{1}{\alpha} \ \! e^{-c_k/\alpha} , \qquad c_k \geq 0 ,
 \label{elaw}
\ee
where the value of $\alpha$ will be determined later on.
It can also be easily shown that this kind of distribution renders the
same results as considering the pulse intensity having the same value,
$C$, for any collision.
In such a case, wherever $\langle c_k \rangle$ and
$\langle c_k^2 \rangle$ appear they have to be replaced
by $C$ and $C^2$, respectively. Finally, we would like to mention that
numerical tests using a different shape function (in particular, a
Gaussian function) gives the same type of results.

Independently of their intensity, it is apparent from Eq.~(\ref{pulse})
that any pulse decays at the same rate $\lambda'$.
This rate defines the decay time scale for collision events,
$\tau_c = 1/\lambda'$.
If $\tau_c$ is relatively small (i.e., collision effects relax
relatively fast), the memory function in Eq.~(\ref{eq-langg}) will
be local in time.
This could be the case, for instance, of relatively dilute systems,
where the average time between consecutive collisions  is long
enough in comparison to the energy transfer process that occurs during
the collision.
Then $\gamma(t-t') \simeq \gamma \delta(t-t')$ and the upper time limit
can be extended to infinity.
In doing so we obtain a standard Langevin equation
\be
 \ddot{x}(t) = - \lambda \dot{x}(t) + F[x(t)] + \delta R(t) .
 \label{eq-lang1}
\ee
Here the friction coefficient $\gamma$ measures the number of
collisions per time unit, as $\lambda$ in Eq.~(\ref{poisson}).
Hence, from now on, we replace $\gamma$ by $\lambda$.
The collisional friction coefficient introduces a time scale
$\tau_r = 1/\lambda$, which can be interpreted as the (average) time
between two successive collisions.
Thus, although each individual collision lasts for a time scale given
by $\tau_c$, its effects over the system take place in a time of the
order of $\tau_r$ in getting damped.

In order to obtain some relevant information about the adsorbate
diffusion process, it is important first to analyze the solutions
of Eq.~(\ref{eq-lang1}) and then to derive some average magnitudes
of interest.
The former are straightforwardly obtained by formal integration,
this yielding
\begin{widetext}
\bs
\ba
 v(t) & = & v_0 e^{- \lambda t}
   + \int_0^t e^{- \lambda (t-t')} F[x(t')] \ \! dt'
   + \int_0^t e^{- \lambda (t-t')} \delta R(t') \ \! dt' ,
 \label{veloc1} \\
 x(t) & = & x_0 + \frac{v_0}{\lambda} \ \! ( 1 - e^{- \lambda t} )
   + \frac{1}{\lambda} \int_0^t \Big[ 1 - e^{- \lambda (t-t')} \Big]
    F[x(t')] \ \! dt'
   + \frac{1}{\lambda} \int_0^t \Big[ 1 - e^{- \lambda (t-t')} \Big]
    \delta R(t') \ \! dt' ,
 \label{posit12}
\ea
 \label{pair}
\es
\end{widetext}
where $v_0 = v(0)$ and $x_0 = x(0)$.
As can be seen, for $\delta R = 0$, Eqs.~(\ref{pair}) are the formal
solution of purely (dissipative) deterministic equations of motion.
Hence, without loss of generality, they can be more conveniently
expressed as
\bs
\ba
 v(t) & = & v_d(t) + v_s(t) ,
 \label{veloc2} \\
 x(t) & = & x_d(t) + x_s(t) ,
 \label{posit2}
\ea
\es
where $d$ embraces the deterministic terms of the solutions and $s$
those other depending on the stochastic force.
Note that when $\delta R \ne 0$ the deterministic part will also
present some stochastic features due to the evaluation of $F(x)$
along $x(t)$.

Since $\langle \delta R(t) \rangle = 0$,
\bs
\ba
 \langle v(t) \rangle & = & \bar{v}_d(t) ,
 \label{avveloc12} \\
 \langle v^2(t) \rangle & = & \bar{v}^2_d(t) + \langle v_s^2(t) \rangle ,
 \label{avveloc22} \\
 \langle x(t) \rangle & = & \bar{x}_d(t) ,
 \label{avposit12} \\
 \langle x^2(t) \rangle & = & \bar{x}^2_d(t) + \langle x_s^2(t) \rangle ,
 \label{avposit22}
\ea
 \label{avvalues2}
\es
where the barred magnitudes indicate the respective averages of the
deterministic part of the solution, and
\bs
\ba
 \langle v_s^2(t) \rangle & = &
  e^{- 2 \lambda t} \int_0^t dt' \ \! e^{2 \lambda t'}
   \int_{-t'}^{t-t'} e^{\lambda \tau} \mathcal{G}(\tau) \ \! d\tau ,
 \label{stocveloc1} \\
 \langle x_s^2(t) \rangle & = &
   \frac{1}{\lambda^2}
   \int_0^t dt' \ \! \Big[ 1 - e^{- \lambda (t-t')} \Big]
  \nonumber \\ & & \times
   \int_{-t'}^{t-t'} \Big[ 1 - e^{- \lambda (t-t'-\tau)} \Big] \ \!
    \mathcal{G}(\tau) \ \! d\tau .
 \label{stocposit1}
\ea
 \label{stocvalues1}
\es
In these expressions, $\mathcal{G}(\tau)$ is the time correlation
function of the stochastic force,
\be
 \mathcal{G}(\tau) \equiv
  \langle \langle \delta R(t) \delta R(t') \rangle \rangle =
  \langle \langle \delta R(t)
   \delta R(t + \tau) \rangle \rangle
 \label{gtau1}
\ee
[the double bracket is defined as in Eq.~(\ref{doubleav})].
As seen in Eq.~(\ref{stocvalues1}), $\mathcal{G}(\tau)$ is the same for
both averages despite the time-dependent prefactors being different.
Averaging over time implies that Eq.~(\ref{gtau1}) is independent of
time; average values and time-correlation functions will not depend
specifically on the origin of time, but on the difference $\tau$
between two times considered.
This condition defines the process as {\it stationary}.

A general expression for $\mathcal{G}(\tau)$ is readily obtained
after straightforward algebraic manipulations to be
\be
 \mathcal{G}(\tau) = \frac{1}{\mathcal{T}} \sum_K P_K (\mathcal{T}) K
  \int_0^\mathcal{T} \langle b(t - t') b(t + \tau - t') \rangle_c
   \ \! dt' ,
 \label{firstg}
\ee
where
\be
 \langle \ \cdot \ \rangle_c \equiv \int_0^\infty \cdot \ g(c) \ \! dc
\ee
is the average over the pulse intensity, with $g(c)$ given by
Eq.~(\ref{elaw}).
Taking into account that $\sum_K P_K (\mathcal{T}) K =
\lambda \mathcal{T}$ and introducing the change of variable
$\zeta = t - t'$, Eq.~(\ref{firstg}) can be approximated by
\be
 \mathcal{G}(\tau) =
  \lambda \int_{-\infty}^\infty
   \langle b(\zeta) b(\zeta + \tau) \rangle_c \ \! d\zeta ,
 \label{corrf}
\ee
which is a general expression independent of the pulse shape.
This approximation relies on the hypothesis that $b(\zeta) \approx 0$
outside a narrow time interval $0 < \zeta < \Delta$, with $\Delta$
a few times larger than $\tau_c$, but of the same order of magnitude.
In particular, substituting Eq.~(\ref{pulse}) into Eq.~(\ref{corrf})
leads to
\be
 \mathcal{G}(\tau) =
  \frac{\lambda \lambda'}{\alpha^2} \ \! {\rm e}^{- \lambda' |\tau|} .
 \label{corrfe}
\ee

By changing the order of the integration variables, integrating over
$t'$, and taking advantage of the property $\mathcal{G}(-\tau) =
\mathcal{G}(\tau)$, Eqs.~(\ref{stocvalues1}) can be expressed as
\begin{widetext}
\bs
\ba
 \langle v_s^2(t) \rangle & = & \frac{2}{\alpha^2} \ \! \bigg\{
  \frac{\lambda'}{2 (\lambda' - \lambda)}
   \Big( 1 - e^{- 2 \lambda t} \Big)
  - \frac{\lambda' \lambda}{\lambda'^2 - \lambda^2}
    \Big[ 1 - e^{- (\lambda' + \lambda) t} \Big] \bigg\} ,
 \label{stocveloce} \\
 \langle x_s^2(t) \rangle & = & \frac{2}{\alpha^2} \ \! \bigg\{
   \frac{t}{\lambda}
  - \frac{2 \lambda' - \lambda}
   {(\lambda' - \lambda) \lambda^2}
     \Big( 1 - e^{- \lambda t} \Big)
  + \frac{\lambda'}{2 (\lambda' - \lambda) \lambda^2}
   \Big( 1 - e^{- 2 \lambda t} \Big)
 \nonumber \\
  & & \qquad + \frac{1}{(\lambda' - \lambda) \lambda'}
     \Big( 1 - e^{- \lambda' t} \Big)
  - \frac{1}{\lambda'^2 - \lambda^2}
   \Big[ 1 - e^{- (\lambda' + \lambda) t} \Big] \bigg\} .
 \label{stocposite}
\ea
 \label{stocvaluese}
\es
\end{widetext}

In order to determine the value of $\alpha$, we assume that
$\lambda' \gg \lambda$, this rendering
\bs
\ba
 \langle v_s^2(t) \rangle & \approx &
  \frac{1}{\ \alpha^2} \Big( 1 - e^{- 2 \lambda t} \Big) ,
 \label{stocveloce2} \\
 \langle x_s^2(t) \rangle & \approx &
  \frac{1}{\ \alpha^2 \lambda^2} \Big[ 2 \lambda t + 1
   - \Big( 2 - e^{- \lambda t} \Big)^2 \Big] .
 \label{stocposite2}
\ea
 \label{stocvaluese2}
\es
Moreover, when $t \to \infty$, we also assume that the equipartition
theorem holds, and therefore
\be
 \frac{1}{2} \ \! m \langle v^2(\infty) \rangle =
  \frac{1}{2} \ \! k_B T .
\ee
Taking into account that $\bar{v}^2_d(t) = \bar{v}^2_0 e^{-2\lambda t}$
and that the time-dependent term in Eq.~(\ref{stocveloce2}) vanish
asymptotically, we obtain $\alpha = \sqrt{m/k_B T}$.
On the other hand, if we consider that the system is initially
thermalized (i.e., it follows a Maxwell-Boltzmann distribution in
velocities) and has a uniform probability distribution in positions
around $x=0$, then $\bar{v}_0 = 0$, $\bar{v}^2_0 = k_B T / m$, and
$\bar{x}_0 = 0$.
Thus, for $\lambda' \gg \lambda$ (i.e., in the Poissonian white noise
limit), Eqs.~(\ref{avvalues2}) become
\bs
\ba
 \langle v(t) \rangle & = & 0 ,
 \label{avveloc13} \\
 \langle v^2(t) \rangle & = & \frac{k_B T}{m} ,
 \label{avveloc23} \\
 \langle x(t) \rangle & = & 0 ,
 \label{avposit13} \\
 \langle x^2(t) \rangle & = & \bar{x}^2_0 +
  \frac{k_B T}{m \lambda^2} \ \!
  \Big[ 2 \lambda t + 1 - \Big( 2 - e^{- \lambda t} \Big)^2 \Big] .
  \nonumber \\ & &
 \label{avposit23}
\ea
 \label{avvalues3}
\es
These equations constitute a limit.
Therefore, for values of the parameters out of the range of the
approximation, deviations are expected.
As will be seen, the behavior of the numerical results presented in
Sec.~\ref{sec4} fit fairly well the trends given by these equations,
though $\lambda'$ is not much larger than $\lambda$ purposedly.
For example, for high values of $\lambda$, the equilibrium thermal
velocity is given by $2f/\alpha^2$, with $f=\lambda'/2(\lambda'+\lambda)$,
instead of $1/\alpha^2$ [note that if $\lambda' = r \lambda$, then
$f = r/2(1 + r)$, which approaches $1/2$ when $r \to \infty$].

After using the approximation $\lambda' \gg \lambda$,
Eqs.~(\ref{avvalues3}) coincide with those obtained for a Brownian
motion.
This is because within this approximation the shot noise behaves as a
(Poissonian) white noise, displaying a behavior analogous to that of
a Gaussian white noise (Brownian motion) whenever the number of
collisions per time unit ($\lambda$) is very high and/or the total
propagation time ($\mathcal{T}$) considered is sufficiently long.
Note that in the limit where the number of collisions goes to infinity,
the Poissonian distribution approaches a Gaussian one because of the
central limit theorem.

As happens for pure Brownian motion ($V = 0$), two regimes are
clearly distinguishable from Eqs.~(\ref{avvalues3}) when comparing
$\lambda$ and $t$.
For $\lambda t \ll 1$, collision events are rare and the adparticle
shows an almost free motion with relatively long mean free paths.
This is the {\it ballistic} or {\it free-diffusion regime},
characterized by
\be
 \langle x^2(t) \rangle \sim \frac{k_B T}{m} \ \! t^2 .
 \label{avvalues4}
\ee
On the other hand, for $\lambda t \gg 1$, there is no free diffusion
since the effects of the stochastic force (collisions) are dominant.
This is the {\it diffusive regime}, where mean square displacements
are linear with time, i.e., they follow {\it Einstein's law},
\be
 \langle x^2(t) \rangle \sim \frac{2 k_B T}{m \lambda} \ \! t = 2 D t .
 \label{avvalues5}
\ee
In analogy to Brownian motion, from this last expression we also note
for systems driven by a shot noise that (1) lowering the density of the
particle gas (or, equivalently, $\lambda$) leads to a faster diffusion
(the diffusion coefficient $D$ increases), and (2) the latter becomes
more active when the gas is heated.

Apart from the averages given above, it is also meaningful to compute
the velocity autocorrelation function,
\be
 \mathcal{C}(\tau) \equiv
 \langle v(0) \ \! v(\tau) \rangle \equiv
  \lim_{\mathcal{T}\to\infty} \frac{1}{\mathcal{T}}
  \int_0^\mathcal{T} v(t) \ \! v(t+\tau) \ \! dt ,
 \label{vcorr1}
\ee
where the correlation time, which provides information about the
line shape broadening (see Sec.~\ref{sec3}), is
\be
 \tilde{\tau} \equiv \frac{1}{\langle v_0^2 \rangle}
  \int_0^\infty \mathcal{C}(\tau) \ \! d\tau .
 \label{tauc}
\ee
The velocity autocorrelation function can also be expressed as
\begin{multline}
 \mathcal{C}(\tau) = \langle v(t) \ \! v(t+\tau) \rangle \\
  = e^{- 2 \lambda t - \lambda \tau} \int_0^t dt' \ \!
    e^{2 \lambda t'}
   \int_{-t'}^{t+\tau-t'} e^{\lambda s} \mathcal{G}(s) \ \! ds .
 \label{vcorr2}
\end{multline}

Introducing Eq.~(\ref{corrfe}) into the right hand side of the second
equality of Eq.~(\ref{vcorr2}), and in the limit of large $t$, we reach
\be
 \mathcal{C}(\tau) =
   \frac{k_B T}{m} \frac{\lambda'^2\lambda}{\lambda'^2 - \lambda^2}
    \left( \frac{e^{- \lambda \tau}}{\lambda} -
           \frac{e^{- \lambda' \tau}}{\lambda'} \right) ,
 \label{corrGM1}
\ee
which in the limit $\lambda' \gg \lambda$ becomes
\be
 \mathcal{C}(\tau) = \frac{k_B T}{m} \ \! e^{- \lambda \tau} .
 \label{corrGM2}
\ee
This expression has been derived starting from a colored shot noise.
However, observe that it is identical to the velocity autocorrelation
function corresponding to an Ornstein-Uhlenbeck process, the
{\it only} stationary Gaussian diffusion process.
This is because of the approximation used here: the number of
collisions per time unit is very large.
In this case, the central limit theorem applies and, according to
{\it Doob's theorem}, the corresponding correlation function will
display an exponential decay.

For $V \ne 0$, Eqs.~(\ref{avvalues3}) are expected to show some
deviations at short time scales because of the role played by the
terms corresponding to the deterministic force.
This force leads to the presence of a confining potential, which makes
the particles to move within a bound space region for a time.
Hence the system will be more localized than in the free-potential
case ($V = 0$) and there will be much less diffusion.
Moreover, some in-phase correlations will also be apparent due to the
particles that do not have energy enough to overcome the potential
energy barrier.
This gives rise to a nondiffusive type of motion.
That is, a low frequency vibrational motion exerted by the so-called
{\it frustrated translational mode} or {\it T mode}.

The velocity autocorrelation functions given by Eqs.~(\ref{corrGM1})
and (\ref{corrGM2}) correspond to a dynamics ruled by completely
flat surface.
Physically, this would be the case of a low corrugated surface, for
which the static force is negligible in both directions.
On the contrary, when the surface corrugation is important, effects
mediated by the $T$ mode are expected to manifest in the correlation
function.
An interesting example to examine is that of a harmonic oscillator,
which can model at a first order of approximation the oscillating
behavior associated to the $T$ mode.
In this case \cite{risken,JLvega1}
\be
 \mathcal{C}(\tau) = \frac{k_B T}{m} \ \! e^{- \lambda \tau/2}
  \left( \cos \omega \tau
   - \frac{\lambda}{2 \omega} \sin \omega \tau \right) ,
 \label{corrHO}
\ee
where
\be
 \omega = \sqrt{ \omega_0^2 - \frac{\lambda^2}{4} }
\ee
and $\omega_0$ is the harmonic frequency; for a nonharmonic potential,
$\omega_0$ represents the corresponding approximate harmonic frequency.
Equation~(\ref{corrHO}) can also be written as
\be
 \mathcal{C}(\tau) =
  \frac{k_B T}{m} \ \! \frac{\omega_0}{\omega} \ \!
   e^{- \lambda \tau/2} \cos (\omega \tau + \delta) ,
 \label{corrHO2}
\ee
with
\be
 \delta \equiv (\tan)^{-1} \left( \frac{\lambda/2}{\omega} \right) .
\ee
Observe that Eq.~(\ref{corrGM2}) is easily recovered in the limit
$\omega_0 \to 0$ from either Eq.~(\ref{corrHO}) or Eq.~(\ref{corrHO2}).

Finally, a brief discussion on the validity of Eq.~(\ref{eq-lang1})
is worthy.
As said above, this equation can only be used rigorously when having
a {\it white noise}, i.e., when $\mathcal{G}(\tau) \sim \delta (\tau)$
and the memory function can be substituted by a constant.
Otherwise, with colored noise, one has to use the generalized Langevin
equation (\ref{eq-langg}) \cite{hanggi2}.
According to the fluctuation-dissipation theorem \cite{kubo}, the
friction in the Langevin equation is related to the fluctuations of
the random force.
This is formally expressed by the relationship between the frequency
spectrum of the memory function and the random force correlation
function,
\be
 \gamma (\omega) = \frac{m}{k_B T}
  \int_0^\infty \mathcal{G}(\tau) \ \! e^{- i \omega \tau} \ \! d\tau .
 \label{fdt}
\ee
Introducing Eq.~(\ref{corrfe}) into Eq.~(\ref{fdt}) yields
\be
 \gamma (\omega) = \lambda \ \! \frac{\lambda'}{\lambda' + i \omega} ,
 \label{fdte}
\ee
whose real part is
\be
 {\rm Re} [ \gamma (\omega) ]
  = \frac{1}{2} \ \! [ \gamma (\omega) + \gamma^* (\omega) ]
  = \lambda \ \! \frac{\lambda'^2}{\lambda'^2 + \omega^2} .
 \label{rfdte}
\ee
Two limits are interesting in this expression: $\lambda' \ll \omega$
and $\lambda' \gg \omega$.
The first limit involves very short time scales (smaller than $\tau_c$),
where the particularities of the adsorbate-adsorbate interaction
potential have to be taken into account (for example, pairwise
interaction), and therefore a MDL prescription has to be followed.
In this case, Eq.~(\ref{fdte}) can be written as
\be
 \gamma (\omega) \approx \lambda \ \! \frac{\lambda'^2}{\omega^2} ,
 \label{limit1}
\ee
and hence the standard Langevin equation can no longer be used.
Moreover, as a consequence, since the frequency depends on the friction
the relaxation time scale $\tau_r$ is not well defined.
Conversely, in the second case, the collision time scale rules the
system dynamics, since it establishes a (frequency) cutoff.
This leads to
\be
 \gamma (\omega) \approx \lambda
  \left( 1 - \frac{\omega^2}{\lambda'^2} \right)
 \label{gammaapprox}
\ee
which can be written as $\gamma (\omega) \sim \lambda$ [notice that
$\lambda \equiv \gamma (0)$] whenever $\lambda \ll \omega \ll \omega_c
= \tau_c^{-1}$.
As can be seen, in this limit (which can also be written as $\tau_c
\ll \tau_r$), the use of a standard Langevin equation is well justified.
This limit holds for strong but localized (or instantaneous) collisions
(as assumed here) as well as for weak but continuous interactions
(Brownian motion).


\section{Elements of surface diffusion}
\label{sec3}

In diffusion experiments carried out by means of quasielastic
helium atom surface scattering (QHAS), one usually measures the
{\it differential reflection coefficient}.
In analogy to liquids \cite{vanHove} this magnitude is given as
\ba
 \frac{d^2 \mathcal{R} (\Delta {\bf K}, \omega)}{d\Omega d\omega}
 & = & n_d \mathcal{F} S(\Delta {\bf K}, \omega)
  \nonumber \\
 & = & n_d \mathcal{F} \iint G({\bf R},t)
  e^{i(\Delta {\bf K} \cdot {\bf R} -\omega t)}
   \ \! d{\bf R} \ \! dt . \nonumber \\ & &
 \label{eq:DRP}
\ea
This expression is the probability that the probe (He) atoms
scattered from the diffusing collective (chattered on the surface)
reach a certain solid angle $\Omega$ with an energy exchange
$\hbar\omega =E_f - E_i$ and wave vector transfer parallel to the
surface $\Delta {\bf K} = {\bf K}_f - {\bf K}_i$.
In Eq.~(\ref{eq:DRP}), $n_d$ is the (diffusing) surface concentration
of adparticles; $\mathcal{F}$ is the {\it atomic form factor}, which
depends on the interaction potential between the probe atoms in the
beam and the adparticles on the surface; and $S(\Delta {\bf K},\omega)$
is the {\it dynamic structure factor} or {\it scattering law}, which
provides a complete information about the dynamics and structure of the
adsorbates through particle distribution functions.
The dynamic structure factor is therefore the observable magnitude
that we are interested in here.
Experimental information about long distance correlations is obtained
from it when using small values of $\Delta {\bf K}$, while information
on long time correlations is available at small energy transfers,
$\hbar \omega$.
On the other hand, a standard procedure employed to obtain the
adiabatic adsorption potential $V({\bf R})$ mediating the
adsorbate-substrate interaction consists in starting with a model
potential that contains some adjustable parameters.
Then, Markovian-Langevin equations are solved for different friction
coefficients with Gaussian white noise to reproduce the experimental
QHAS measurements \cite{toennies2}.

When dealing with interacting particles, particle distribution
functions are described by means of the so-called van Hove or
time-dependent pair correlation function $G({\bf R},t)$
\cite{vanHove}.
Given a particle at the origin at some arbitrary initial time,
$G({\bf R},t)$ represents the average probability for finding the
same or another particle at the surface position ${\bf R}=(x,y)$
at time $t$.
This function thus generalizes the well known pair distribution
function $g({\bf R})$ from statistical mechanics
\cite{mcquarrie,Hansen}, since it provides information about the
interacting particle dynamics.
Depending on whether we consider correlations of a particle with
itself or with other different particles, we distinguish between
{\it self} correlation functions, $G_s({\bf R},t)$, and {\it distinct}
correlation functions, $G_d({\bf R},t)$, respectively.
With this, the full (classical) pair correlation function can be
expressed as
\be
 G({\bf R},t) = G_s({\bf R},t) + G_d({\bf R},t) .
 \label{gtotal}
\ee
According to its definition, $G_s({\bf R},t)$ is peaked at
\mbox{$t=0$}, and approaches zero as $t$ increases since the particle
loses correlation with itself.
On the other hand, at $t = 0$, $G_d({\bf R},t)$ gives the static pair
correlation function (the standard pair distribution function),
$g({\bf R}) \equiv G_d({\bf R},0)$, while approaches the mean surface
number density $\sigma$ of diffusing particles as $t \to \infty$.
Taking this into account, Eq.~(\ref{gtotal}) can be expressed as
\be
 G({\bf R},0) = \delta ({\bf R}) + g({\bf R})
 \label{gtotal0}
\ee
at $t = 0$, and as
\be
 G({\bf R},t) \approx \sigma
 \label{gtotalinfty}
\ee
for a homogeneous system with $\|{\bf R}\| \to \infty$ and/or
$t \to \infty$. At low adparticle concentrations, when
interactions among adsorbates can be neglected because they are
far apart from each other, the main contribution to Eq.~(\ref{gtotal})
is $G_s$ (particle-particle correlations are negligible and
$G_d \approx 0$).
On the contrary, for high coverages, it is expected that $G_d$
contributes significantly to Eq.~(\ref{gtotal}).
Within our approach, the interaction among adsorbates is described
by a particle subjected to a random force (a shot noise).
Thus, diffusion is described by the $G = G_s$ function in
the {\it interacting single adsorbate approximation}. At low
coverages, the $G_s$ function in both approximations (interacting and
noninteracting) has to be the same.

As seen in Sec.~\ref{sec2}, although Gaussian white noise arising from
the surface is not considered here, an analogous Markovian-Langevin
also emerges from our model based on shot noise.
Therefore the same analytical results obtained elsewhere \cite{JLvega1}
with a Gaussian white noise can be easily extended to our case, where
the friction coefficient has to be interpreted in terms of the
collision frequency between adsorbates.
For this treatment, the dynamic structure factor is better expressed as
\cite{vanHove}
\ba
 S(\Delta {\bf K},\omega)& = &
  \int e^{-i\omega t} \ \!
   \langle e^{-i\Delta {\bf K} \cdot {\bf R}(t)}
    e^{i\Delta {\bf K} \cdot {\bf R}(0)}\rangle \ \! dt
  \nonumber \\
   & = & \int e^{-i\omega t} \ \! I(\Delta{\bf K},t) \ \! dt ,
 \label{eq:DSF}
\ea
where the brackets in the integral denote an ensemble average and
${\bf R}(t)$ is the adparticle trajectory.
Here,
\be
 I(\Delta {\bf K},t) \equiv
  \langle e^{-i\Delta {\bf K} \cdot
   [{\bf R}(t) - {\bf R}(0)] } \rangle
  = \langle e^{-i \Delta K \int_0^t v_{\bf K} (t') \ \! dt'} \rangle
 \label{eq:IntSF}
\ee
is the {\it intermediate scattering function}, which is the space
Fourier transform of $G({\bf R},t)$ \cite{note1}.
In Eq.~(\ref{eq:IntSF}), $v_{\Delta {\bf K}}$ is the velocity of the
adparticle projected onto the direction of the parallel momentum
transfer, with length $\Delta K \equiv \| \Delta {\bf K} \|$.
After a second-order cumulant expansion in $\Delta K$ in the second
equality of Eq.~(\ref{eq:IntSF}), the intermediate scattering function
reads as
\be
 I(\Delta {\bf K},t) \approx
  e^{- \Delta K^2 \int_0^t (t - t') \mathcal{C}(t') dt'} ,
 \label{eq:IntSF2}
\ee
where $\mathcal{C}(t)$ is given by Eq.~(\ref{vcorr1}) [note that
$\mathcal{C}(t)$, as given by Eq.~(\ref{vcorr1}), is independent
of the direction of $\Delta {\bf K}$].
This is the so-called {\it Gaussian approximation} \cite{mcquarrie},
which is exact when the velocity correlations at more than two
different times are negligible.
This allows one to replace the average acting over the exponential
function in Eq.~(\ref{eq:IntSF}) by an average acting over its
argument, as seen in Eq.~(\ref{eq:IntSF2}).

In the case of an almost flat surface the resulting intermediate
scattering function is
\be
 I(\Delta {\bf K},t) = \exp \left[- \chi^2
   \left( e^{- \lambda t} + \lambda t - 1 \right) \right] ,
 \label{eq:IntSGM}
\ee
with
\be
 \chi^2 \equiv \langle v_0^2 \rangle \Delta K^2 / \lambda^2.
\ee
From this relation we can obtain both the mean free path, $\bar{l}
\equiv \tau_r \sqrt{ \langle v_0^2 \rangle }$, and the self-diffusion
coefficient, $D \equiv \tau_r \langle v_0^2 \rangle$ (Einstein
relation).
It can be easily shown that the dynamic structure factor derived from
Eq.~(\ref{eq:IntSGM}) has a Gaussian shape for short times compared to
$\tilde{\tau}=\tau_r$ (or $\chi \rightarrow \infty$) and a Lorentzian
shape for long times or values of small $\Delta K$ (or $\chi \ll 1$)
\cite{JLvega1,jpcm2}.
In this last case, Eq.~(\ref{eq:IntSGM}) becomes a pure decaying
exponential function,
\be
 I(\Delta {\bf K},t) \approx e^{- \chi^2 \lambda t} ,
 \label{eq:IntSGMapprox}
\ee
which is in agreement with the ansatz used in Ref.~\cite{jardine2}
to fit the polarization values obtained experimentally.

On the other hand, when a harmonic oscillator is considered,
Eq.~(\ref{eq:IntSF2}) becomes
\be
 I(\Delta {\bf K},t) =
  \exp \left\{ - \frac{\chi^2 \lambda^2}{\omega\omega_0}
 \left[ \cos \delta - e^{-\lambda t/2} \cos (\omega t - \delta) \right]
  \right\} .
 \label{isfho}
\ee
As infers from this equation, the intermediate scattering function
displays an oscillatory, but exponentially damped, behavior around a
certain value [although in the limit $\omega_0 \to 0$ Eq.~(\ref{isfho})
approaches Eq.~(\ref{eq:IntSGM})].
Unlike the free-potential case, this means that after relaxation
some correlation is still present in the system, being the limit
value
\be
 I(\Delta {\bf K},\infty) \to
  e^{- \chi^2 \lambda^2 \cos \delta / \omega \omega_0} .
 \label{isfhoapprox}
\ee
Again, this is in agreement with the experimental observation.
According to Ref.~\cite{jardine2} the residual or anomalous
value of the polarization observed seems to be connected to the
blockage of the adsorbate perpendicular motion (which is related
to translational hopping).
On the other hand, according to our model, another source for such
an effect could be that some Na atoms would become trapped inside
potential wells, thus behaving as damped oscillators.
Nonetheless, note that in both cases the anomalous behavior for $I(t)$
arises from inhibiting the free diffusion of the atoms on the surface.
For realistic corrugated surfaces, one will find a nonseparable
combination of this behavior (related to the trapped particles inside
the potential wells) and that described by Eq.~(\ref{eq:IntSGM})
(associated to the free-diffusing particles).
The corresponding scattering law has been given elsewhere
\cite{JLvega1} for models where the velocity autocorrelation function
can be expressed in terms of a damped anharmonic oscillator.

Taking into account the approximation for the intermediate scattering
function, we can now relate in an easy manner the velocity power
spectrum with the dynamic structure factor.
As can be readily seen from Eq.~(\ref{eq:IntSF2}) \cite{Hansen}, we
have
\be
 \mathcal{C}(t) = - \lim_{\Delta K \to 0}
  \frac{1}{\Delta K^2} \frac{d^2 I(\Delta{\bf K},t)}{dt^2} ,
 \label{eq:ISF}
\ee
where the velocity is recorded along the $\Delta {\bf K}$ direction.
The velocity power spectrum or Fourier transform of the velocity
autocorrelation function is defined as
\be
 Z(\omega) = \int_{-\infty}^\infty
  \mathcal{C}(t) \ \! e^{-i\omega t} \ \! dt .
 \label{eq:PSD}
\ee
Thus, by Fourier transforming both sides in Eq.~(\ref{eq:ISF}) we
obtain
\be
 Z(\omega) = \omega^2 \lim_{\Delta K \to 0}
  \frac{S (\Delta K, \omega)}{\Delta K^2} .
\label{eq:PS-DSF}
\ee
The interest in this expression relies on the fact that it allows us
to obtain a relationship between the diffusion coefficient, $D$, and
the full width at half maximum (FWHM) of $S$ by setting $\omega = 0$
in Eq.~(\ref{eq:PS-DSF}), since
\be
 D = \frac{1}{2} \ \! Z(\omega = 0) .
 \label{eq:DC}
\ee
For example, for a Lorentzian-shaped $S$, the FWHM is given by
$\Gamma = 2 D \Delta K^2$, which is basically the line shape that one
can observe in an experiment.
On the other hand, if the line shape is Gaussian, its FWHM is
$\Gamma = 2 \sqrt{2\ln 2} \ \! \langle v_0^2 \rangle^{1/2}
\ \! \Delta K$.

Before concluding this section, we would like to give a brief
account on how to relate in a simple manner the coverage $\theta$
and $\lambda$.
In the elementary kinetic theory of transport in gases (see, for
example, Ref.~\cite{mcquarrie}), diffusion is proportional to
the mean free path $\bar{l}$, which is proportionally inverse to both
the density of gas particles and the effective area of collision when
a hard-sphere model is assumed.
For 2D collisions, the effective area is replaced by an effective length
(twice the radius $\rho$ of the adparticle) and the gas density by the
surface density $\sigma$.
With this, the mean free path reads as
\be
 \bar{l} = \frac{1}{2 \sqrt{2} \rho \sigma} .
 \label{mfp}
\ee
According to the Chapman-Enskog theory for hard spheres, the
self-diffusion coefficient can be written as
\be
 D = \frac{1}{6 \rho \sigma} \ \! \sqrt{\frac{k_B T}{m}} .
 \label{d-mfp}
\ee
Now, from the Einstein relation, and taking into account that
$\theta = a^2 \sigma$ for a square surface lattice of unit cell
length $a$, we obtain
\be
 \lambda = \frac{6 \rho \ \! \theta}{a^2} \ \! \sqrt{\frac{k_B T}{m}} .
 \label{theta}
\ee
Therefore, given a certain surface coverage and temperature, $\lambda$
can be readily estimated from Eq.~(\ref{theta}).


\section{Results and discussion}
\label{sec4}


\subsection{Numerical details}
\label{sec4.1}

In order to solve Eq.~(\ref{eq-lang1}) we have used the velocity
Verlet algorithm, which is commonly applied when dealing with
stochastic differential equations \cite{allen}.
For the average calculations shown here a number of
$10\ 000$$-$$20\ 000$ trajectories is sufficient for convergence.
The initial conditions are chosen such that the velocities are
distributed according to a Maxwell-Boltzmann velocity distribution
at a temperature $T$, and the positions follow a uniform random
distribution along the extension of a single unit cell of the
potential model used (see below).
Regarding the dynamical parameters, we have used $\lambda' =
10^{-3}$~a.u.\ ($\approx 41.3$~ps$^{-1}$) and for the mass and
radius of the adparticles we have considered those of a Na atom,
since Na on Cu(001) is a typical system where diffusion has been
studied and will serve to illustrate our model numerically.
As for the coverage, $\theta_{\rm Na} = 1$ corresponds to one
Na atom per Cu(001) surface atom or, equivalently,
$\sigma = 1.53 \times 10^{19}$~atom/cm$^2$ \cite{toennies3};
$a = 2.557$~\AA\ is the unit cell length; and $\rho = 2$~\AA\
has been used for the atomic radius.
For example, with these values a collision friction
$\lambda = 5 \times 10^{-6}$~a.u.\ ($\approx 0.2067$~ps$^{-1}$) at
a surface temperature $T = 100$~K would be related to a coverage
$\theta_{\rm Na} = 0.059$, while the same friction at a temperature
$T = 200$~K would be caused by a coverage $\theta_{\rm Na} = 0.042$.
One order of magnitude higher for $\lambda$ implies also one order of
magnitude for the coverage at the corresponding surface temperatures.


\subsection{Diffusion in a flat surface}
\label{sec4.2}

\begin{figure}
 \includegraphics[width=7cm]{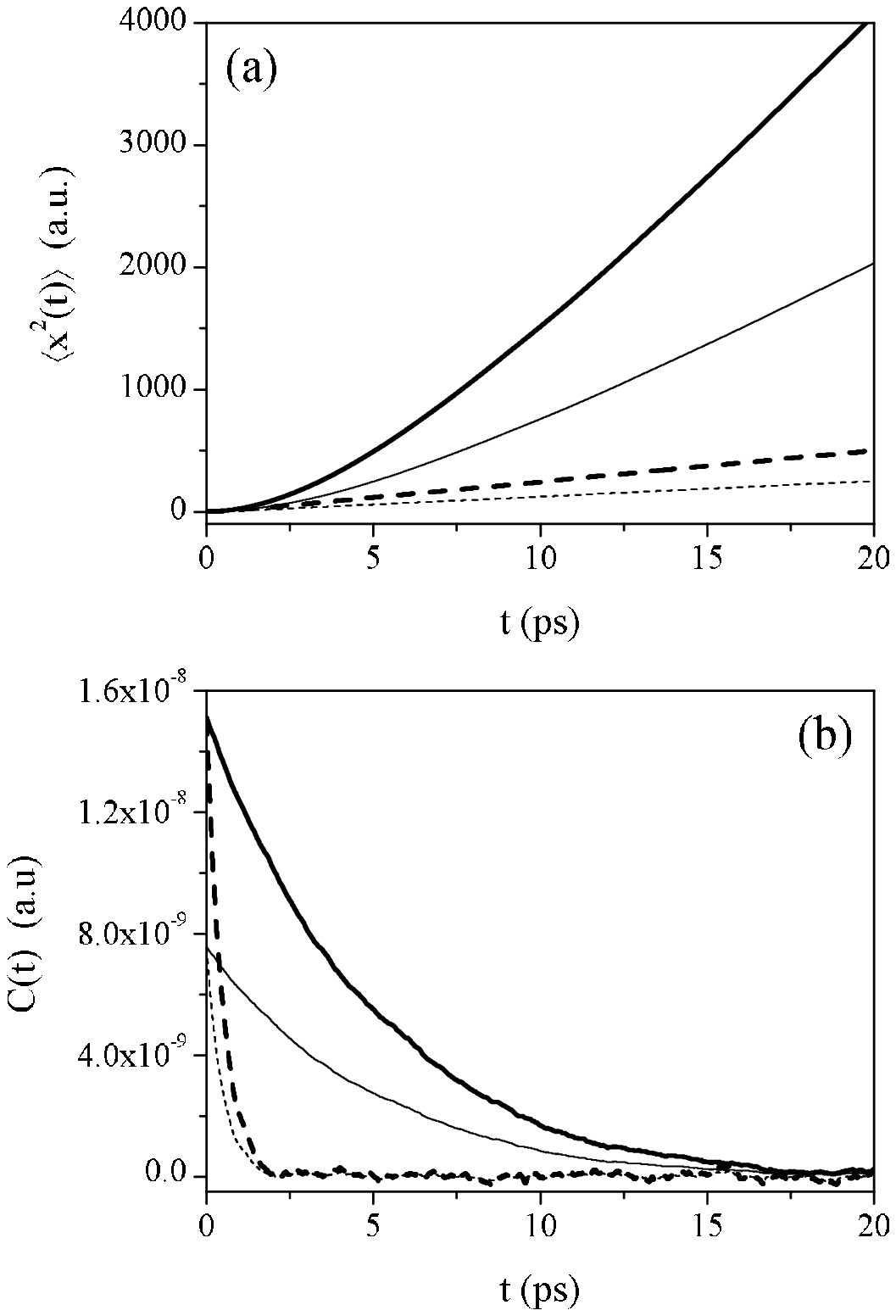}
 \caption{\label{fig1}
  (a) $\langle x^2 (t) \rangle$ for two different temperatures and
  two values of the friction coefficient.
  (b) $\mathcal{C}(t)$ for the same temperatures and friction
  coefficients as in (a).
  The temperatures used are $T = 100$~K (thin lines) and $T = 200$~K
  (thick lines), and the friction coefficients are
  $\lambda = 5\times10^{-6}$~a.u.\ (solid lines) and
  $\lambda = 5\times10^{-5}$~a.u.\ (dashed lines).}
\end{figure}

First we are going to analyze the case of a flat surface, which
represents fairly well the situation of a low corrugated real surface,
where the role of the activation barrier is negligible.
Thus, in Fig.~\ref{fig1}(a) we observe that, effectively, as predicted
by the theory, two (time) regimes are distinguishable from the mean
square displacement $\langle x^2(t) \rangle$: parabolic and linear,
corresponding to free and diffusive motion, respectively.
As is apparent, diffusion (proportional to the slope of the linear
regime) is enhanced by increasing temperature and decreasing the
friction coefficient.
Within the diffusive regime, though the presence of many particles
tends to inhibit individual particle motions, there is still the
possibility for particles to propagate but at a smaller rate [compare
the cuadratic increase of $\langle x^2(t) \rangle$ during the free
motion with its linear increase during the diffusive one].
The effect of the diffusion process can also be seen by looking at
the velocity autocorrelation function $\mathcal{C}(t)$ plotted in
Fig.~\ref{fig1}(b).
As seen, the larger $\lambda$ the faster the decay, this result being
in agreement with Eq.~(\ref{corrGM2}) as well as its exponential decay.
The numerical results fit perfectly on this decay, but with a slightly
lower $\lambda$ than the nominal one because, strictly speaking, ours
is not a rigorous white shot noise.
The agreement can also be observed when increasing the temperature,
leaving $\lambda$ unchanged; the starting value of $\mathcal{C}(t)$
increases, but its trend, $e^{-\lambda t}$, remains the same.

\begin{figure}
 \includegraphics[width=6.4cm]{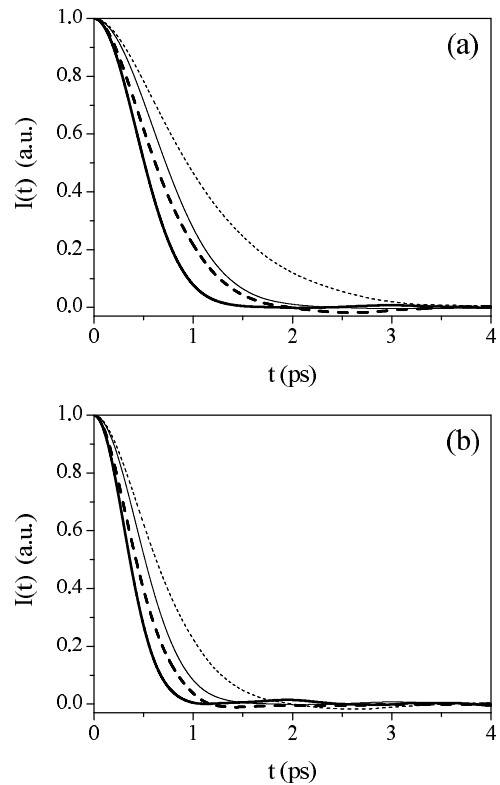}
 \caption{\label{fig2}
  (a) $I(t)$ for two different temperatures and two values of the
  friction coefficient, and $\Delta K = 0.88$~\AA$^{-1}$.
  (b) Same as (a), but for $\Delta K = 1.23$~\AA$^{-1}$.
  The temperatures used are $T = 100$~K (thin lines) and $T = 200$~K
  (thick lines), and the friction coefficients are
  $\lambda = 5\times10^{-6}$~a.u.\ (solid lines) and
  $\lambda = 5\times10^{-5}$~a.u.\ (dashed lines).}
\end{figure}

\begin{figure}
 \includegraphics[width=6.45cm]{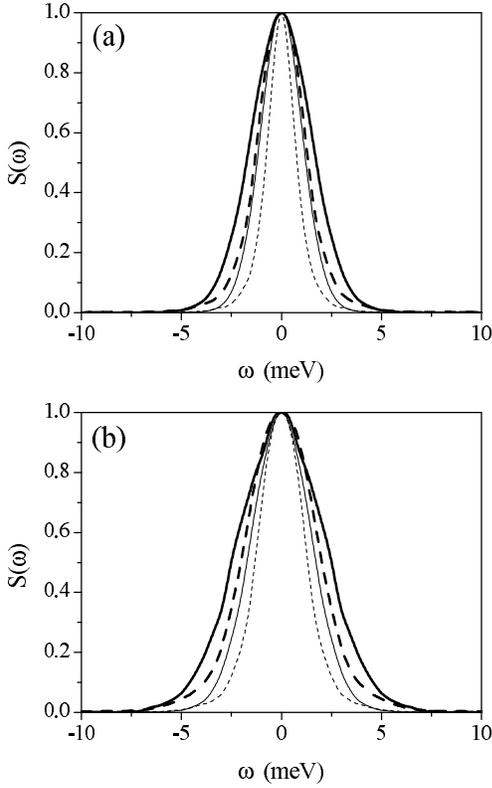}
 \caption{\label{fig3}
  (a) $S(\omega)$ for two different temperatures and two values of
  the friction coefficient, and $\Delta K = 0.88$~\AA$^{-1}$.
  (b) Same as (a), but for $\Delta K = 1.23$~\AA$^{-1}$.
  The temperatures used are $T = 100$~K (thin lines) and $T = 200$~K
  (thick lines), and the friction coefficients are
  $\lambda = 5\times10^{-6}$~a.u.\ (solid lines) and
  $\lambda = 5\times10^{-5}$~a.u.\ (dashed lines).
  In all cases, $S(\omega)$ has been normalized to unity in order to
  better appreciate the line shape broadening/narrowing; moreover, they
  have also been smoothed out to eliminate contributions coming from
  long-time fluctuations in $I(t)$.}
\end{figure}

For a flat surface the Gaussian approximation assumed in
Eq.~(\ref{eq:IntSF2}) for $I(t)$ is exact.
This is numerically corroborated by looking at the results presented
in Fig.~\ref{fig2}, where two different values of $\Delta K$,
0.88~\AA$^{-1}$ and 1.23~\AA$^{-1}$, are analyzed.
As is apparent, for both $\Delta K$ values, the initial falloff of
$I(t)$ displays a Gaussian shape, while for longer times it is
exponential.
This makes that the dynamic structure factor (see Fig.~\ref{fig3}) at
small positive (annihilation events) and negative (creation events)
energy transfers, i.e., the quasielastic peak region, displays a
mixed Gaussian-Lorentzian profile describable by using the $\Gamma$
and incomplete $\Gamma$ functions \cite{JLvega1,jpcm2}.
Notice from these plots that as $\lambda$ increases the
line shape undergoes narrowing, unlike what one would expect when having
collision events.
This result for a flat surface, which is also predicted analytically,
can be explained as follows.
Noninteracting adsorbates behave like an ideal gas, i.e., particles
spread out freely, without feeling the action of any other particle.
This gives rise to dynamic structure factors that display Gaussian
profiles.
However, as particle-particle interactions are taken into account
there is a friction arising from the neighboring adsorbates that
opposes the free motion, which increases with increasing $\lambda$.
That is, going back to the results presented in Fig.~\ref{fig1}(a),
the particle reaches the diffusive regime faster as $\lambda$ becomes
larger. The corresponding line shapes are no longer Gaussian functions
and display narrowing at larger values of $\lambda$ (see
Fig.~\ref{fig3}), as analytically predicted by the scattering law
\cite{JLvega1,jpcm2}.
Of course, as temperature increases, for a fixed $\lambda$, one
evidently observes the broadening of the line shape.

\begin{figure}
 \includegraphics[width=7cm]{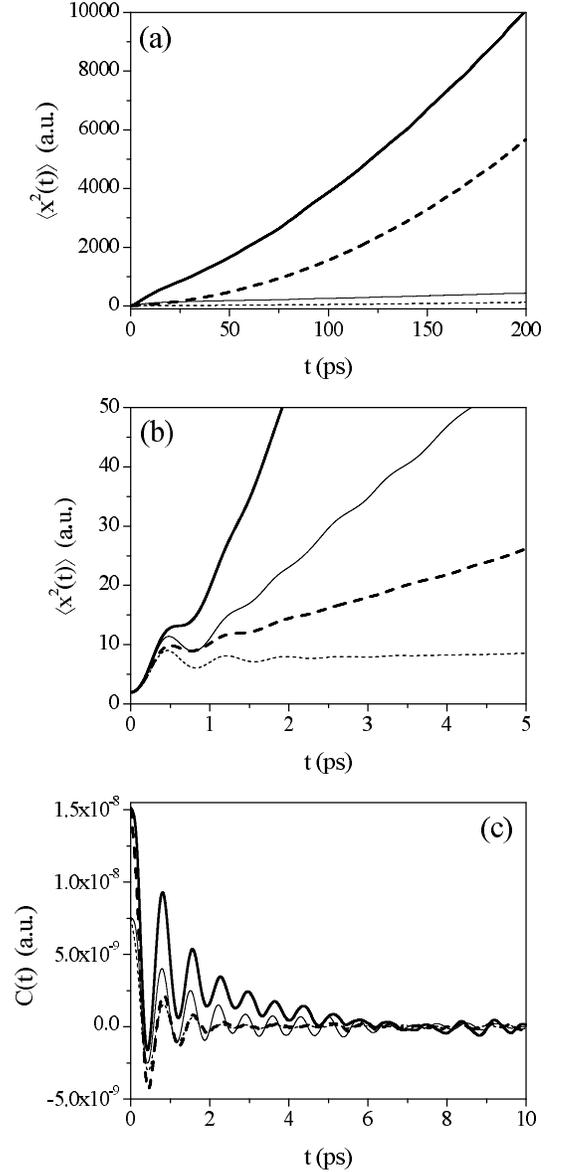}
 \caption{\label{fig4}
  (a) $\langle x^2 (t) \rangle$ for two different temperatures and
  two values of the friction coefficient.
  (b) Enlargement at short times of part (a).
  (c) $\mathcal{C}(t)$ for the same temperatures and friction
  coefficients as in (a).
  The temperatures used are $T = 100$~K (thin lines) and $T = 200$~K
  (thick lines), and the friction coefficients are
  $\lambda = 5\times10^{-6}$~a.u.\ (solid lines) and
  $\lambda = 5\times10^{-5}$~a.u.\ (dashed lines).}
\end{figure}

Finally, observe that the results presented here are similar to those
that one would obtain with a Brownian-like motion: (i) with $\lambda$ a
slow diffusion is observed according to the Einstein relation, and
(ii) diffusion becomes more active as the temperature of the ensemble
of adparticles increases.
However, also note that unlike a Brownian motion, diffusion appears
here as a consequence of the discrete (in time) ``kicks'' felt by
the particles.
Moreover, between two consecutive ``kicks'' or collisions, they move
basically without feeling any stochastic force (this is similar to
the collision model proposed in Ref.~\cite{tommei}).


\subsection{Diffusion in a separable potential}
\label{sec4.3}

\begin{figure}
 \includegraphics[width=6.4cm]{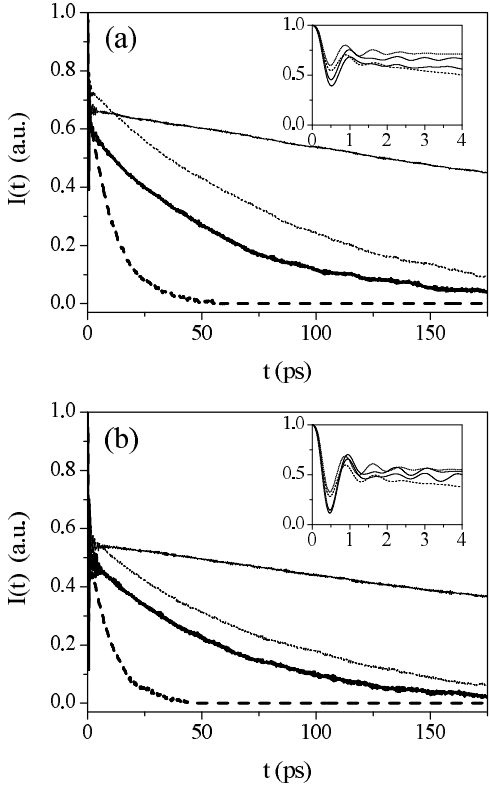}
 \caption{\label{fig5}
  (a) $I(t)$ for two different temperatures and two values of the
  friction coefficient, and $\Delta K = 0.88$~\AA$^{-1}$.
  (b) Same as (a), but for $\Delta K = 1.23$~\AA$^{-1}$.
  The temperatures used are: $T = 100$~K (thin lines) and $T = 200$~K
  (thick lines); and the friction coefficients are:
  $\lambda = 5\times10^{-6}$~a.u.\ (solid lines) and
  $\lambda = 5\times10^{-5}$~a.u.\ (dashed lines).}
\end{figure}

When the surface corrugation is relatively strong and cannot be
neglected, it will induce very important effects regarding the
adsorbate dynamics.
To illustrate these effects here we consider an adsorbate-surface
interaction potential model
\be
 V(x,y) = V_0 [2 - \cos (2 \pi x/a) - \cos (2 \pi y/a)] ,
 \label{pot}
\ee
where $2 V_0 = 33.5$~meV is the activation barrier height in one
direction ($x$ or $y$).
As a function of the temperature, the behavior is the same as
previously observed for a flat surface.
As can be seen in Fig.~\ref{fig4}(a) [and in an enlargement
view in Fig.~\ref{fig4}(b)], for a given temperature the presence of
the potential also inhibits the diffusion of particles at larger values
of $\lambda$ since the number of them trapped in the potential wells is
also larger (according to Einstein's law).
This result is also reported in Refs.~\cite{toennies3,ying} when the
coverage increases.
However, the effect of the static interaction potential will give
rise to observing the opposite behavior than in the case of a flat
surface: as $\lambda$ increases line shapes become broader, a result
which is also observed experimentally \cite{toennies3}.
This is because now both motions, diffusion and vibration, are coupled.
Thus, with increasing $\lambda$, the adsorbates can remain localized
inside a given surface well for longer times.
This allows one to observe a higher collisional damping in the $T$
mode, which participates actively in the diffusion process.
This is clearly depicted in Fig.~\ref{fig4}(c), where the velocity
autocorrelation function displays an oscillatory behavior, which is
more damped as $\lambda$ increases.
We observe that $\mathcal{C}(t)$ fits the profile given by
Eq.~(\ref{corrHO2}) but with parameters different from a
harmonic oscillator (see Ref.~\cite{JLvega1} for an anharmonic
oscillator model).
Only for a motion localized mainly at the bottom of the potential
well, the harmonic oscillator profile will be reproduced.

\begin{figure}
 \includegraphics[width=6.45cm]{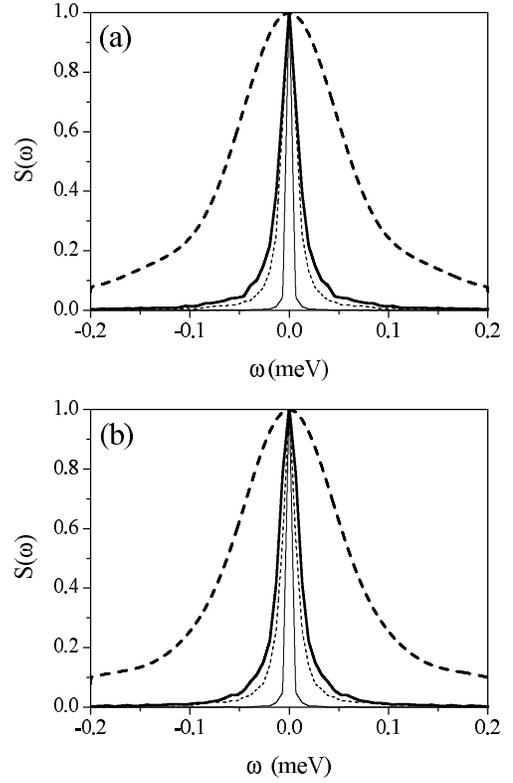}
 \caption{\label{fig6}
  (a) $S(\omega)$ for two different temperatures and two values of
  the friction coefficient, and $\Delta K = 0.88$~\AA$^{-1}$.
  (b) Same as (a), but for $\Delta K = 1.23$~\AA$^{-1}$.
  The temperatures used are: $T = 100$~K (thin lines) and $T = 200$~K
  (thick lines); and the friction coefficients are:
  $\lambda = 5\times10^{-6}$~a.u.\ (solid lines) and
  $\lambda = 5\times10^{-5}$~a.u.\ (dashed lines).
  In all cases, $S(\omega)$ has been normalized to unity in order to
  better appreciate the line shape broadening or narrowing; moreover,
  they have also been smoothed out to eliminate contributions coming
  from long-time fluctuations in $I(t)$.}
\end{figure}

The oscillations due to the $T$ mode are not only observable in a
plot of $\mathcal{C}(t)$, but they also manifest in $I(t)$
(see Fig.~\ref{fig5}) and $S(\omega)$ (see Fig.~\ref{fig6}) for two
values of $\Delta K$, 0.88~\AA$^{-1}$ and 1.23~\AA$^{-1}$, covering
the region of the first Brillouin zone, $[0,1.3]$~\AA$^{-1}$.
As shown elsewhere \cite{JLvega1}, the FWHM of the quasielastic peak
is quite similar for both $\Delta K$ values and therefore the variation
obtained in the width is due only to the effect of $\lambda$.
Plotting $I(t)$ we can observe a loss of phase when comparing this
function for different values of $\lambda$.
This behavior continues for all oscillations and ends up with $I(t)$
falling faster for the case with larger $\lambda$, thus given rise
to broadening as $\lambda$ increases.
Nonetheless, it is expected that for relatively large values of
$\lambda$, one can recover the behavior observed in the flat case.

\begin{figure}
 \includegraphics[width=6.45cm]{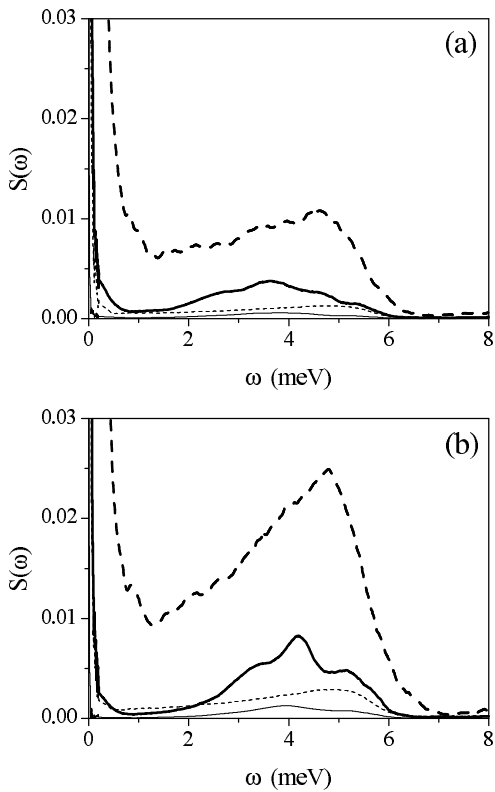}
 \caption{\label{fig7}
  Same as Fig.~\ref{fig6}, but enlarged to show the peaks corresponding
  to the $T$ mode.
  Since this mode is symmetric with respect to $\omega = 0$~meV, only
  the one located to the right of the quasielastic peak is shown in
  every case.}
\end{figure}

Finally, in Fig.~\ref{fig7} the peak corresponding to the $T$ mode
placed around the frequency of oscillation (4~meV in energy) is
plotted for the same two values of $\Delta K$ as before.
Not only a shift of the position but also a broadening are clearly
seen with $\lambda$, as also observed experimentally \cite{toennies3}.
For both the quasielastic peak and the $T$ mode peak the analytical
formulas for the line shapes given elsewhere \cite{JLvega1} in the
context of Gaussian white noise also fit fairly well the numerical
results presented for the white shot noise model used here.

\section{Conclusions}
\label{sec5}

We would like to stress that the same treatment developed here
for a separable 2D periodic potential can be easily generalized to a
nonseparable 2D periodic surface with substrate friction due to a
Gaussian white noise \cite{prb}, this model describing the process in
a more realistic fashion. In particular, preliminary results with
the nonseparable 2D periodic surface for a low coverage
($\theta_{\rm Na} = 0.028$) have rendered
a value $\Gamma = 120$~$\mu$eV for the quasielastic peak width at
$\Delta K = 1.26$~\AA$^{-1}$, which is in agreement with the
experimental and theoretical values, $\Gamma_{\rm exp} = 110$~$\mu$eV
and $\Gamma_{\rm th} = 110$~$\mu$eV, respectively, given in
Ref.~\cite{toennies3}.

The broadening obtained by this shot noise model agrees qualitatively
well with the QHAS experimental observations \cite{toennies3}.
Note that the broadening described by our model is of the same type
as that observed in the spectral lines of gases under high pressure
conditions.
Moreover, in our simple model the important issue of how the presence
of more and more adatoms changes the fields of force felt by the
remaining adparticles is not addressed.
As far as we know, this problem has not been treated in the literature
and certainly merits further investigation.
It would be very interesting to know how it affects the broadening
of the quasielastic peak.
Nevertheless, at the level of the experiments carried out it could
happen that the statistical limit (a large number of collisions during
the time scales considered) would not lead to any relevant feature,
since any effect linked to type of interaction would be blurred up
with time.

Regarding the validity of our model, since it is purely stochastic,
in principle there are no other limitations than those imposed by
realistic physical conditions.
For instance, it is clear that the maximum coverage should be
$\theta_{\rm Na} = 1$.
However, as the coverage increases, it is also apparent that
adsorbate-adsorbate interactions will play a more prominent role in
the diffusion dynamics, since in average the motion of each adsorbate
will be slowed down and they will feel for longer times the force
exerted by their neighbors.
In such a case, the Markovian approximation will break down and memory
effects should be taken into account.
On the other hand, according to Ref.~\cite{jardine2} there is
experimental evidence that for coverages greater than $\theta_{\rm Na}
= 0.05$, motion perpendicular to the surface is observed, this
contributing to the diffusion dynamics.
Therefore, one might think that a stochastic model such as ours could
at least work fine up to values of the coverage about $\theta_{\rm Na} =
0.2$ or greater depending on whether QHAS measurements still display
effective Lorentzian functions for quasielastic line shapes
\cite{toennies3}.


\section*{Acknowledgments}

This work was supported in part by DGCYT (Spain) under project
No.~FIS2004-02461.
R.M.-C.\ would like to acknowledge the University of Bochum for support
from the Deutsche Forschungsgemeinschaft, Contract No.~SFB 558.
J.L.V.\ and A.S.S.\ would like to acknowledge the Spanish Ministry of
Education and Science for a grant and a ``Juan de la Cierva'' contract,
respectively.
The authors also thank Dr.\ A.\ Jardine for interesting discussions and
comments.



\begin{thebibliography}{99}
\bibitem{coverage}
 The surface coverage, $\theta$, is defined as the number of adsorbed
 molecules on a surface divided by the number of molecules in a filled
 monolayer on that surface.
\end{thebibliography}
\end{document}